\documentclass{article}

\usepackage[utf8]{inputenc}
\usepackage{amsmath}
\usepackage{amssymb}
\usepackage{amsthm}
\usepackage{xparse}
\usepackage{mathtools}
\usepackage{tikz}
\usepackage{subfig}
\usepackage{hyperref}
\usepackage[capitalize]{cleveref}

\newtheorem{theorem}{Theorem}

\DeclareDocumentCommand\transpose{m}{#1^{\intercal}}
\DeclareDocumentCommand\Z{}{\mathbb{Z}}
\DeclareDocumentCommand\R{}{\mathbb{R}}
\DeclareDocumentCommand\Znonneg{}{\Z_{\geq 0}}
\DeclareDocumentCommand\conv{o}{\operatorname{conv}\IfValueTF{#1}{\left(#1\right)}{}}
\DeclareDocumentCommand\cplxNP{}{\mathsf{NP}}
\DeclareDocumentCommand\cplxcoNP{}{\mathsf{coNP}}

\title{The Graphical Traveling Salesperson Problem has no Integer Programming Formulation in the Original Space}
\author{Matthias Walter}
\date{\today}

\begin{document}

\maketitle

\begin{abstract}
  The Graphical Traveling Salesperson Problem (GTSP) is the problem of assigning, for a given weighted graph, a nonnegative number $x_e$ to each edge $e$ such that the induced multi-subgraph is of minimum weight among those that are spanning, connected and Eulerian.
  Naturally, known mixed-integer programming formulations use integer variables $x_e$ in addition to others.
  Denis Naddef posed the challenge of finding a (reasonably simple) mixed-integer programming formulation that has integrality constraints only on these edge variables.
  Recently, Carr and Simonetti (IPCO 2021) showed that such a formulation cannot consist of polynomial-time certifyiable inequality classes unless $\cplxNP = \cplxcoNP$.
  In this note we establish a more rigorous result, namely that no such MIP formulation exists at all.
\end{abstract}

\section{Introduction}

Let $G = (V,E)$ be a graph and let $c \in \mathbb{R}^E$.
The Graphical Traveling Salesperson problem is about finding $c$-minimum cost tour in $G$ that visits each node at least once, where edges can be used multiple times.
It can be formulated as the following constraint integer program due to Corn{\'e}jols, Fonlupt and Naddef~\cite{CornuejolsFN85}.
\begin{subequations}
  \label{cip_graphical_tsp}
  \begin{alignat}{7}
    & \text{min } ~\mathrlap{\transpose{c}x} \label{cip_graphical_tsp_obj} \\
    & \text{s.t. }
      & \sum_{e \in \delta(S)} x_e &\geq 2 &\quad& \forall \varnothing \neq S \subsetneqq V \label{cip_graphical_tsp_connectivity} \\
    & & \sum_{e \in \delta(v)} x_e & ~\text{is even} &\quad& \forall v \in V \label{cip_graphical_tsp_parity} \\
    & & x_e &\in \Znonneg &\quad& \forall e \in E \label{cip_graphical_tsp_x}
  \end{alignat}
\end{subequations}
Here, $\delta(S) \coloneqq \{ e \in E : |e \cap S| = 1 \}$ and $\delta(v) \coloneqq \delta(\{v\})$ denote the cuts induced by node set $S \subseteq V$ and node $v \in V$, respectively.
For each edge $e \in E$, the variable $x_e$ indicates how often $e$ is traversed in the tour.

\DeclareDocumentCommand\Pgtsp{}{P_{\mathop{gtsp}}}

The authors of~\cite{CornuejolsFN85} describe several classes of inequalities that are valid for the \emph{GTSP polyhedron} $\Pgtsp(G)$ defined as the convex hull of all feasible solutions, i.e.,
\begin{align*}
  \Pgtsp(G) \coloneqq \conv\{ x \in \Z^E : x \text{ satisfies~\eqref{cip_graphical_tsp_connectivity}, \eqref{cip_graphical_tsp_parity} and~\eqref{cip_graphical_tsp_x}} \}.
\end{align*}
Among these were \emph{path}, \emph{wheelbarrow} and \emph{bicycle} inequalities.
In order to turn~\eqref{cip_graphical_tsp} into a mixed-integer programming model (MIP), constraint~\eqref{cip_graphical_tsp_parity} can be replaced by this pair of constraints:
\begin{subequations}
  \begin{align}
    \sum_{e \in \delta(v)} x_e &= 2y_v &\quad& \forall v \in V \\
    y_v &\in \Z &\quad& \forall v \in V
  \end{align}
\end{subequations}
These additional $y$-variables are artificial and their presence has no impact on the linear programming relaxation of~\eqref{cip_graphical_tsp}.
For this reason, Naddef posed the challenge of finding a simple mixed-integer programming formulation that involves, apart from the $x$-variables, only continuous variables~\cite{Naddef19}.
According to~\cite{CarrS21}, he had the formulation from~\cite{CornuejolsFN85} with \emph{path}, \emph{wheelbarrow} and \emph{bicycle} inequalities in mind.
There exist other (mixed-)integer programming formulations for the GTSP, see~\cite{CarrS21,ErmelW20}, all of which requiring additional \emph{integral} variables.

Recently, Carr and Simonetti showed that such a formulation cannot be nice in the sense that it cannot consist of inequality families for which one can certify membership in polynomial time, provided $\cplxNP \neq \cplxcoNP$ (see Section~4.2 in~\cite{CarrS21}).
The purpose of this paper is to show that the reason for the non-existence of a simple formulation does not lie in complexity theory.
In fact, we show that no such formulation exists at all:

\begin{theorem}
  \label{thm_graphical_tsp_no_mip}
  The GTSP has no mixed-integer programming formulation whose only integer variables are the $x$-variables from~\eqref{cip_graphical_tsp}.
\end{theorem}

\section{Nonexistence of the formulation}

\begin{figure}[htb]
  \begin{center}
    \subfloat[Instance $G^\star$ with unit costs $c^\star$]{
      \label{fig_example_instance}
      \begin{tikzpicture}[
        node/.style={very thick, draw=black, circle, inner sep=0mm, minimum size=4mm},
        edge/.style={very thick, draw=black}
      ]
        \node[node] (s) at (0,0) {$s$};
        \node[node] (t) at (2,0) {$t$};
        \node[node] (a) at (1,0.7) {$a$};
        \node[node] (b) at (1,0) {$b$};
        \node[node] (c) at (1,-0.7) {$c$};
        \draw[edge] (s) to[bend left] node[anchor=south, font=\footnotesize] {$1$} (a);
        \draw[edge] (a) to[bend left] node[anchor=south, font=\footnotesize] {$1$} (t);
        \draw[edge] (s) -- node[anchor=south, font=\footnotesize] {$1$} (b) -- node[anchor=south, font=\footnotesize] {$1$} (t);
        \draw[edge] (s) to[bend right] node[anchor=north, font=\footnotesize] {$1$} (c);
        \draw[edge] (c) to[bend right] node[anchor=north, font=\footnotesize] {$1$} (t);
      \end{tikzpicture}
    }
    \subfloat[First tour $x^\star_1$]{
      \label{fig_example_tour1}
      \begin{tikzpicture}[
        node/.style={very thick, draw=black, circle, inner sep=0mm, minimum size=4mm},
        edge/.style={very thick, draw=black}
      ]
        \node[node] (s) at (0,0) {$s$};
        \node[node] (t) at (2,0) {$t$};
        \node[node] (a) at (1,0.7) {$a$};
        \node[node] (b) at (1,0) {$b$};
        \node[node] (c) at (1,-0.7) {$c$};
        \draw[edge] (s) to[bend left] (a);
        \draw[edge] (a) to[bend left] (t);
        \draw[edge] (s) to[bend left=15] (b);
        \draw[edge] (s) to[bend right=15] (b);
        \draw[edge] (s) to[bend right] (c);
        \draw[edge] (c) to[bend right] (t);
      \end{tikzpicture}
    }
    \subfloat[Second tour $x^\star_2$]{
      \label{fig_example_tour2}
      \begin{tikzpicture}[
        node/.style={very thick, draw=black, circle, inner sep=0mm, minimum size=4mm},
        edge/.style={very thick, draw=black}
      ]
        \node[node] (s) at (0,0) {$s$};
        \node[node] (t) at (2,0) {$t$};
        \node[node] (a) at (1,0.7) {$a$};
        \node[node] (b) at (1,0) {$b$};
        \node[node] (c) at (1,-0.7) {$c$};
        \draw[edge] (s) to[bend left] (a);
        \draw[edge] (a) to[bend left] (t);
        \draw[edge] (b) to[bend left=15] (t);
        \draw[edge] (b) to[bend right=15] (t);
        \draw[edge] (s) to[bend right] (c);
        \draw[edge] (c) to[bend right] (t);
      \end{tikzpicture}
    }
  \end{center}
  \caption{%
    Example instance $G^\star = (V^\star,E^\star)$ with unit costs $c^\star = \transpose{(1,1,1,1,1,1)}$.
    To minimum-cost solutions are depicted in~\eqref{fig_example_tour1} and~\eqref{fig_example_tour2}.
  }
  \label{fig_example}
\end{figure}

\begin{proof}[Proof of \cref{thm_graphical_tsp_no_mip}]
  We consider the graph $G^\star = (V^\star, E^\star)$ from \cref{fig_example} with unit edge costs $c^\star \in \R^{E^\star}$.
  It is easy to see that $G^\star$ has no Hamiltonian cycle, and therefore there is no solution of value $|V^\star| = 5$.
  Hence, the tours in Figure~\eqref{fig_example} \subref{fig_example_tour1} and~\subref{fig_example_tour2}, denoted by $x^\star_1, x^\star_2 \in \Z^{E^\star}$ are optimal.
  Their midpoint is the point $x^\star = \transpose{(1,1,1,1,1,1)}$, which is integral but infeasible for~\eqref{cip_graphical_tsp} as it violates~\eqref{cip_graphical_tsp_parity}.

  Assume, for the sake contradiction, that there exists a mixed-integer programming formulation $Q = \{ (x,y) \in \Z^E \times \R^q : Ax + By \leq d \}$ that has integrality constraints only for the $x$-variables.
  Hence, the projection of $Q$ onto the $x$-variables is the set of feasible solutions to~\eqref{cip_graphical_tsp}, and hence
  \begin{equation}
    \min \{ \transpose{c}x : (x,y) \in Q \} \label{mip_graphical_tsp}
  \end{equation}
  is equivalent to~\eqref{cip_graphical_tsp}.
  In particular, feasibility of $x^\star_1$ and $x^\star_2$ for~\eqref{cip_graphical_tsp} implies that there exist $y^\star_1, y^\star_2 \in \R^q$ such that $(x^\star_i,y^\star_i) \in Q$ for $i=1,2$.
  Now let $y^\star \in \R^q$ be the midpoint of $y^\star_1$ and $y^\star_2$.
  By convexity of the linear relaxation of $Q$ and integrality of $x^\star$, also $(x^\star,y^\star)$ is an optimal solution to~\eqref{mip_graphical_tsp}.
  This contradicts the fact that $x^\star$ is infeasible for~\eqref{cip_graphical_tsp}.
\end{proof}

\paragraph{Acknowledgments.}
We thank R.\ Carr and N.\ Simonetti for stimulating discussions on the challenge of Naddef as well as an anonymous referee whose comments led to an improved presentation of the material.

\bibliographystyle{plain}
\bibliography{graphical-tsp-no-ip}

\end{document}